\documentclass[a4paper,11pt]{article}
\usepackage{jinstpub} 
\usepackage{lineno}

\proceeding{LIDINE2022 - Light Detection In Noble Elements\\
 21-23 September 2022\\
 University of Warsaw Library}

\title{\boldmath Lindhard integral equation with binding energy applied to
 light and charge yields of nuclear recoils in noble liquid
 detectors}

\author{Y. Sarkis, Aguilar-Arevalo, and Juan Carlos D'Olivo}
\affiliation{Instituto de Ciencias Nucleares, Universidad Nacional Autónoma de México, 04510 CDMX, Mexico}

\emailAdd{youssef@ciencias.unam.mx}

\abstract{We present a model of the ionization efficiency, or quenching factor, for low-energy
nuclear recoils, based on a solution to Lindhard integral equation with binding energy and apply it
to the calculation of the relative scintillation efficiency and charge yield for nuclear recoils in noble
liquid detectors. The quenching model incorporates a constant average binding energy together with
an electronic stopping power proportional to the ion velocity, and is an essential input in an analysis
of charge recombination processes to predict the ionization and scintillation yields. Our results
are comparable to NEST simulations of LXe and LAr and are in good agreement with available
data. These studies are relevant to current and future experiments using noble liquids as targets for
neutrino physics and the direct searches for dark matter.}


\arxivnumber{10.1088/1748-0221/18/03/C03006} 

\begin{document}
\maketitle
\flushbottom
\section{Introduction}
\label{sec:intro}
Noble liquid time projection chambers (TPCs) employing liquid argon or xenon (LAr or LXe) as detection medium are a powerful technology for the measurement of the low energy nuclear recoils (NR) expected from the direct detection of dark matter \cite{Zeplin3_2009,XENON1T2018,Darkside2018,PandaX2017,LUX2017,PANDAX-II2020} and the coherent elastic neutrino nucleus scattering (CE$\nu$NS). In these detectors the signals coming from ionization and scintillation  are typically used to discriminate NR from other signals. Proper modeling  the production of charge and light quanta by NR in such detectors is key to understanding their response, specially at sub-keV energies, where interesting hints for new physics may lie. In particular, CE$\nu$NS, \footnote{ recently observed by the COHERENT collaboration in a CsI crystal and in LAr \cite{CEVNSDetcoherent,CEVNSLAr}} can open a new channel for the study of different processes beyond the standard model.
Scintillation and charge production  from nuclear recoils are quenched relative to electronic recoils of the same energy \cite{BasicsQFTPC}. The nuclear stopping power (nuclear scattering) becomes relevant compare to electronic stopping (ionization)\cite{lindhard:1963} at energies below $\lesssim 1$ MeV. Direct dark matter searches and CE$\nu$NS experiments  are sensitive  to the combination of the quenching of nuclear recoil light and charge yield, and the low energy recoil signature.
In 1963, Lindhard \cite{lindhard:1963} gave a model\footnote{$f_n=kg(\varepsilon)/(1+kg(\varepsilon)),\quad g(\varepsilon)=\varepsilon +3\varepsilon^{0.15} +6\varepsilon^{0.7}$}  that works only for nuclear recoil energies  $\gtrsim 10$ keV, since his model is just an approximate solution to the equations, where the atomic binding energy was neglected. Hence, the application of this model  for most of the direct dark matter and CE$\nu$NS experiments is in principle not justified because of the aforementioned arguments. Recently, it was shown that the exact numerical solution of the integral equation  that incorporate the atomic binding energy \cite{QFUNAM2020} describe properly the ionization energy given by a nuclear recoil at low energies in Si and Ge. In this work, we are going to study the application of this recent approach to LAr and LXe.     
\section{Energy Dissipation in Noble Liquids}
In  noble  detectors, the average ionization energy   $\mathcal{\bar H}$  that a particle with non zero cross section gives to the material target by direct nuclear impact  is divided in three channels: the excitation of intrinsic atoms $N_{ex}$, the formation of electron-ion pairs $N_{i}$, and the average kinetic energy $\tilde{E}$ of sub-excitation electrons that goes into heat. Thus, energy conservation is given by Platzman \cite{PLATZMAN1961116} equation
\begin{equation}\label{Eq:platzman}
{E_R}=(N_{ex})(E_{ex}) + (N_{i})(E_{i}) + N_i\tilde{E},    
\end{equation}
where $E_{R}$ is the total recoil energy given by a particle to the ensemble of atoms, $E_{ex}$ is the average energy expenditure for excited states and $E_{i}$ is the average energy expenditure to create an electron-ion pair. 
For example, in a dual-phase detector
under an electric field,  nuclear recoil energy deposition produces photons (scintillation) and electrons (ionization).
The number of scintillation
photons, $n_{\gamma}$ , is proportional to the amount of ionization plus some
fraction $r$ (with $0 < r < 1$) of charge which recombines with free
ions: $n_\gamma = N_{ex} + rN_i$. Correspondingly for ionization electrons, $n_e = (1-r)N_i$; the damping effect from recombination leads to an anti-correlation between charge and light signals. Some of the electrons generated through ionization in the
liquid are drifted to the gas. Therefore, two signals are measured. The first is the primary
scintillation light due to direct excitation and recombination of e-ion pairs in liquid denoted
as $S_1$ . The second is the proportional scintillation light in gas, denoted as $S_2$, which is proportional to the number of electrons escaping the recombination process. 
We assume that one excitation corresponds to one photon and also that the exciton to ion quotient $N_{ex}/N_i=\beta$ will behave as an effective constant to all recoil energies considered. Accordingly,  the total quanta is
\begin{equation}\label{Eq1}
N=N_{ex}+N_i=n_e+n_{\gamma}=N_i(1+\beta).
\end{equation}  
Meanwhile the total recoil energy $E_R$ given to the liquid is divided mainly into the average ionization energy $\mathcal{\bar H}$ and average atomic motion energy $\mathcal{\bar N}$, i.e.  $E_R=\mathcal{\bar H}+\mathcal{\bar N}$. When atomic binding energy $U$ is taken into account, considering both the  energy needed to set free a moving ion in the media and to excite inner degrees of freedom of the ion, the kinetic energy of the ion at each collision is $E=E_R-U$. In this work, we assume a constant effective binding energy for all recoil energies considered and, we defined ionization efficiency as  
\begin{equation}\label{Eq2}
f_n=\mathcal{\bar H}/E_R,
\end{equation}
where $f_n=0$ if the kinetic energy of the ion is not greater than $U$.i.e, $E\leq U$. By using $f_n$ we can estimate the total quanta in \ref{Eq1} as a function of the recoil energy and the mean energy $W_i$ to create an ion electron pair  in the material, where $W_i=\mathcal{\bar H}/N_i$. Hence 
\begin{equation}\label{Eq3}
\mathcal{\bar H}=W_iN_i=W_iN(1+\beta)^{-1}=W_sN,\; \Rightarrow E_R=W_s(n_e+n_{\gamma})/f_n,
\end{equation} 
where Eq. (\ref{Eq1}) was applied and $W_s=W_i/(1+\beta)$  is the mean energy to create  a ionization  or an excitation in the liquid. For LXe, $W_i=15.6 \pm 0.3$ eV, $W_s=13.7\pm 0.2$ \cite{WiLXe,WiLXeMathew}, and for LAr, $W_i=23.6 \pm 0.4$ eV, $W_s=19.5\pm 1.0$ eV \cite{WiLAr}. For the gas phase in Xe, $W^{g}_i=22.0$ eV and  $W^{g}_i=26.4$ eV in Ar \cite{RecombinationModel}.
\section{Charge Recombination  Model }
The total given  energy  to electrons is divided between the   energy  that goes to $N_{ex}$  excited atoms and  $N_i$  ionized atoms, $\mathcal{\bar H}=(N_{ex})(E_{ex}) + (N_{i})(E_{i})$. By assuming   that
each excited atom in LXe or LAr creates one
scintillation photon, and that each ionized atom produces
a single electron, unless it recombines\cite{Doke_2002}, we have $n_\gamma+n_e=N_{ex} +N_i$. Since the ionization process occurs in a media where ions move freely (ionized gas) surrounded by a liquid (solid) phase, we expect that the effective energy $W^*_i$ to create an electron-pair is between the values in the liquid  and gas  phase $W_i<W^*_i<W^g_i$.     
The recombination fraction $r$ can be modeled by using Thomas-Imel approach \cite{ThomasImel}, 
\begin{equation}\label{Eq4}
 r=1-\frac{1}{\xi} \ln \left(1+\xi\right),\quad \xi=\frac{N_{i} \alpha}{4 a^{2} v}=\gamma N_i \;,
\end{equation}
where ${N_i}={\mathcal{\bar H}}/{W^*_i}={E_R{f_n}}/{W^*_i}$ and the parameter $\gamma= { \alpha}/{4 a^{2} v}$ can depend in the electric field, but here it will be taken as an effective constant \cite{LArFit,WiLXeMathew}. Hence,  the charge and light yields after recombination are
\begin{eqnarray}
n_e=\frac{N_i}{\xi} \ln (1+\xi), \quad n_\gamma=N_i(\beta +1 - \frac{1}{\xi} \ln (1+\xi)).
\end{eqnarray} 
Ionization efficiency $f_n$ is needed  since   $\xi$ and $N_i$ depends on it. 
To model $f_n$ we  use the approach given in \cite{QFUNAM2020}, where the ionization efficiency is computed by solving a second order  integro-differential equation for the average atomic motion $\mathcal{\bar N}$ or $\bar\nu$ in dimensionless units\footnote{$\varepsilon=CE$, where $C=11.5/Z^{7/3}\rm{1/keV}$. } , that includes the effect of the constant binding energy $u$, an electronic stopping power $S_e=k\varepsilon^{1/2}$ \cite{SeFactor},  
\begin{eqnarray}\label{Eq:ModSimIntEq}
k\;\varepsilon^{1/2}\bar\nu'(\varepsilon) -\tfrac{1}{2}k\;\varepsilon^{3/2}\bar\nu''(\varepsilon)  =  
\int^{\varepsilon^{2}}_{\varepsilon u} dt \frac{f(t^{1/2})}{2t^{3/2}}\times\left[\bar\nu(\varepsilon-t/\varepsilon) + \bar\nu(t/\varepsilon -u)-\bar\nu(\varepsilon) \right] .
\end{eqnarray}
Where $f$ is a function proportional to the elastic nuclear stopping power and can be calculated semi-empirically \cite{Ziegler}. The ionization efficiency $f_n$ can be compute directly by $f_n=(\varepsilon+u-\bar\nu)/(\varepsilon+u)$ 
hence the model predicts a threshold at $\varepsilon_R=2u$, since if $\varepsilon=u$  the right hand side of Eq.(\ref{Eq:ModSimIntEq}) is zero (the ion doesn't have enough energy to disrupt the atomic binding). Hence, a discontinuity in the derivatives is present at this point, that can be studied by using the following parametrization,
\begin{equation}
    \bar{\nu}(\varepsilon) = 
    \left\lbrace
      \begin{array}{ll}
         \varepsilon + u       & ,\;\;\varepsilon < u \,\, ,\\
          \varepsilon + u -\lambda(\varepsilon)  & ,\;\;\varepsilon \geq u  \,\, ,
      \end{array}
    \right.
    \label{eq:splitsol}
\end{equation}
where $\lambda$ is a continuous function fulfilling $\lambda(u)=0$ and with discontinuities in the derivatives ($\lambda'(u^+)=\alpha_1$ and ($\lambda''(u^+)=\alpha_2$). We compute $\alpha_2$  to calculate $\bar\nu$ accordingly to Eqs.(\ref{eq:splitsol},\ref{Eq:ModSimIntEq}), having the relation  $\alpha_1 = 1 + (1/2)u\alpha_2$ at $\varepsilon=u$, by using a shooting method that pursues the boundary condition $\bar\nu''(\infty )\rightarrow 0^-$. For details about the algorithm see \cite{QFUNAM2020}. Improvements for considering a more appropriate electronic stopping power can be seen in \cite{BEZRUKOV2011119}. \\
 Here we compare our results with those of the NEST simulation. NEST is a global MC framework, independent of experiments and detectors, that enables
the simulation of scintillation and ionization yield averages and total quanta as functions of deposited energy \cite{Szydagis_2011}. For total quanta in noble liquids
NEST considers a model with $N_q=\alpha E^\beta$ , with $\alpha=11^{+2}_{-0.5}$  $\beta=1.1\pm0.05$ for LXe and, $\alpha=11.1\pm 1.4$ and $\beta=1.087\pm0.01$ for LAr. The ionization and light yield recombination models used by NEST are presented in \cite{NESTData,Szydagis_2011}.
\section{Results}
\begin{figure}[h]
\centering
\includegraphics[scale=0.33]{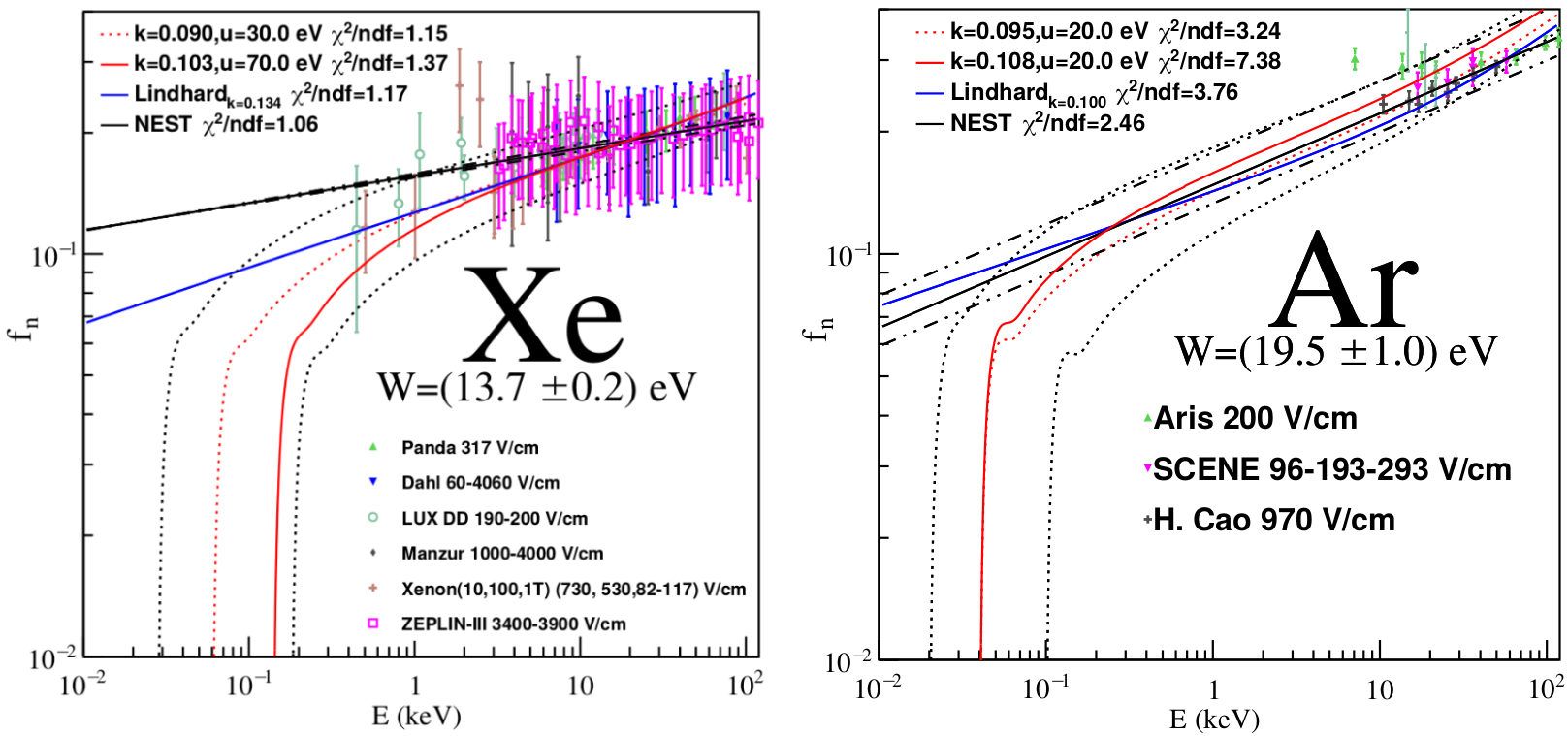}
\caption{Total quanta for LXe and LAr as a function of the recoil energy. Measurements are taken from \cite{NESTData}. The black dotted lines are  the approximate cover of measurements error bars, red dotted line is the best fit of the model  to the data, solid red line is the model used to compute light and charge yields, solid blue line  is the Lindhard's model and solid black is the NEST curve with error curves draw line dotted black.} \label{Fig1}
\end{figure}

\begin{figure}
\includegraphics[scale=0.39]{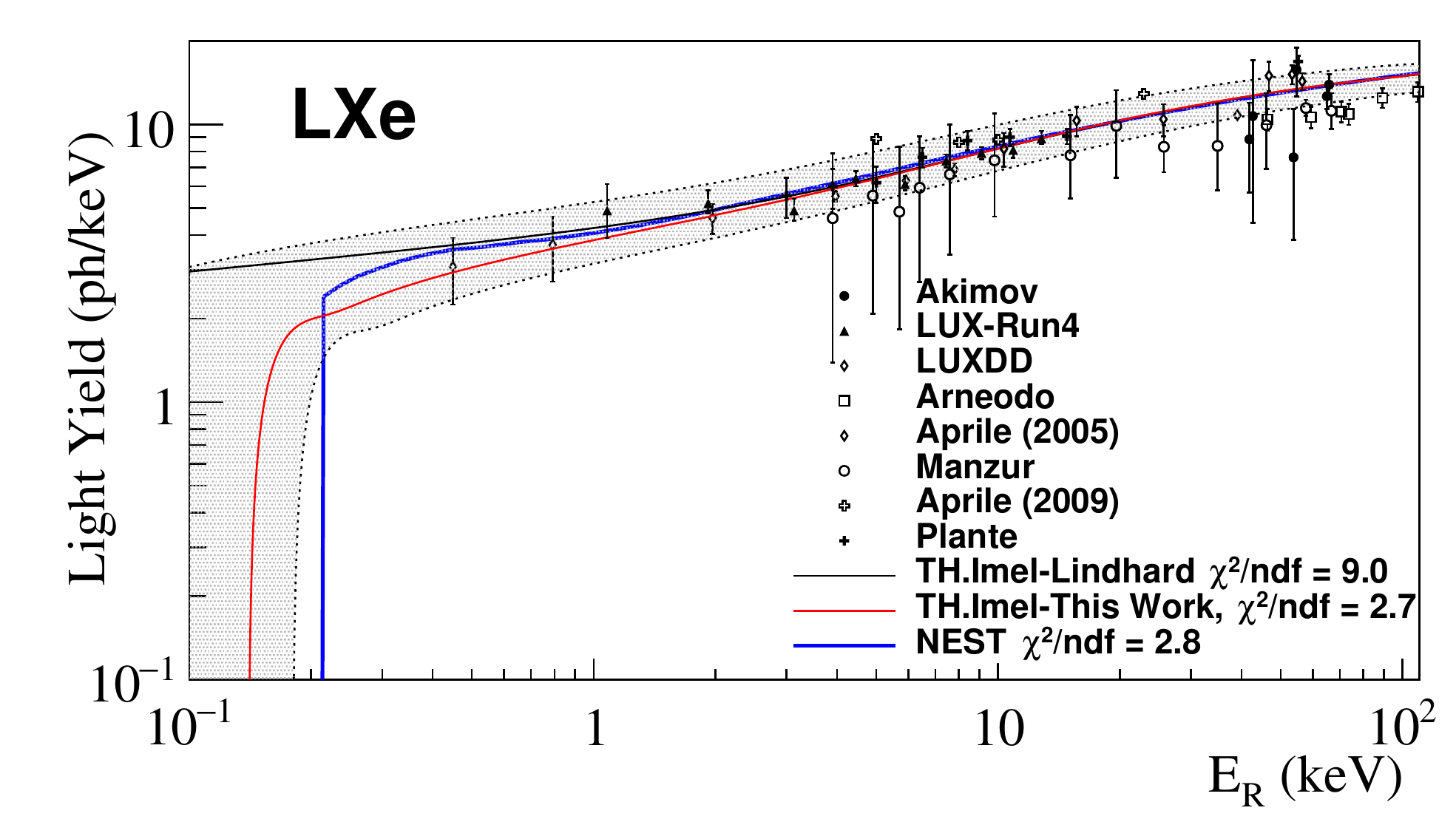}
\includegraphics[scale=0.39]{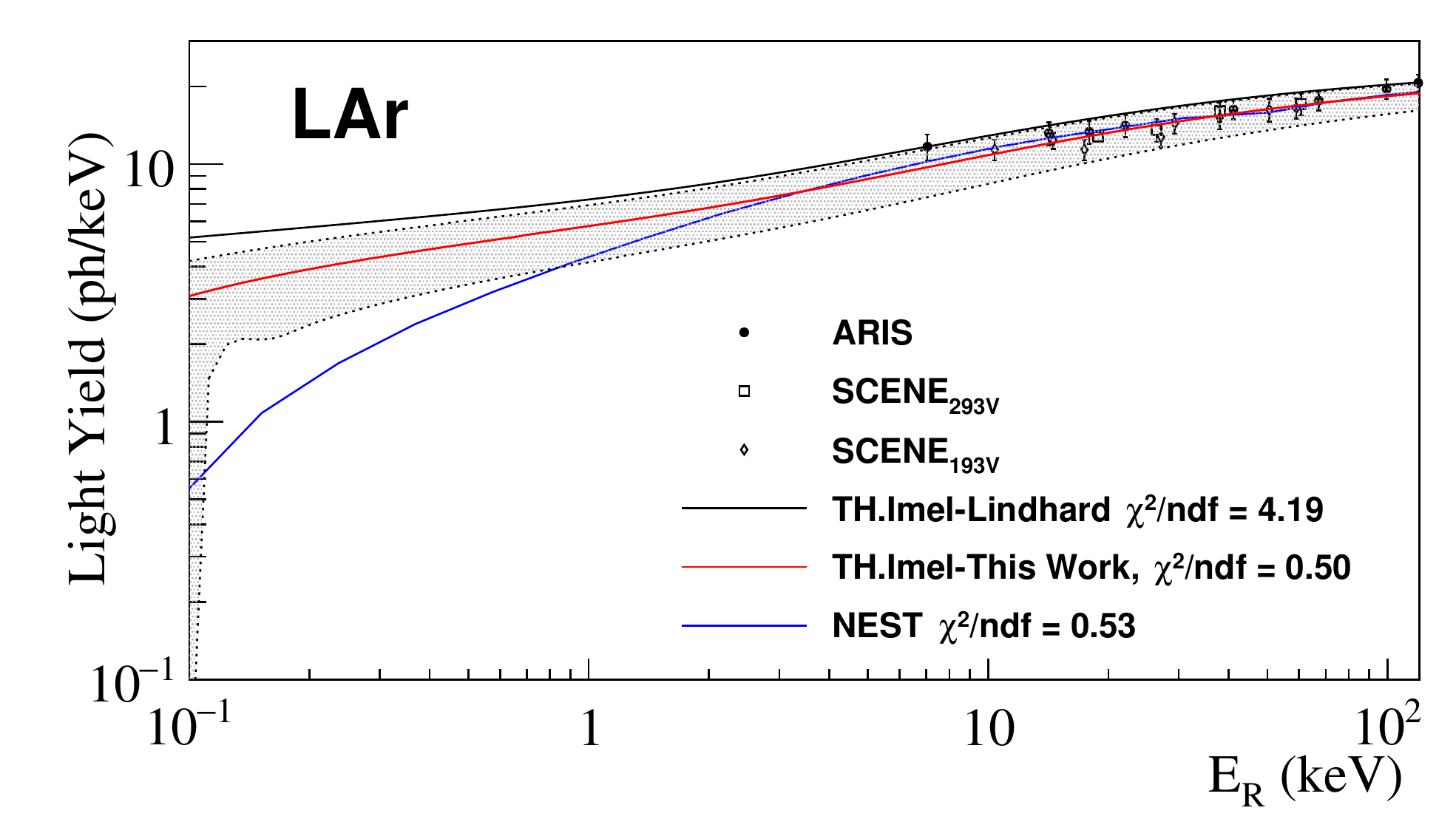}
\caption{Light yield for LXe and LAr   as a function of the recoil energy. The solid red line is the best fit  using Thomas-Imel box model with the ionization efficiency computed according to \cite{QFUNAM2020}, solid black is Thomas-Imel box model with Lindhard's ionization efficiency and solid blue is NEST curve. }\label{Fig2}
\end{figure}

\begin{figure}
\includegraphics[scale=0.39]{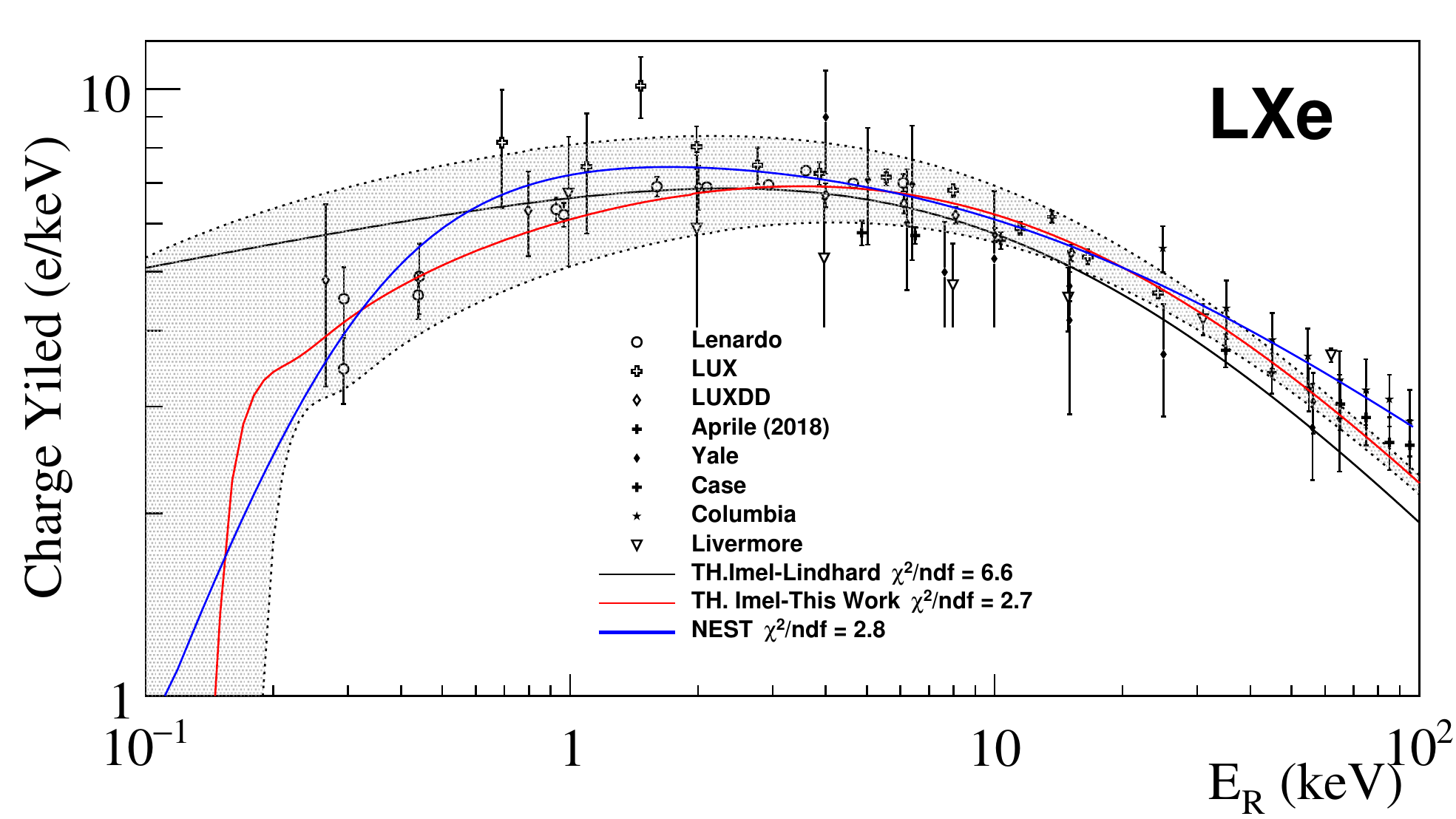}
\includegraphics[scale=0.39]{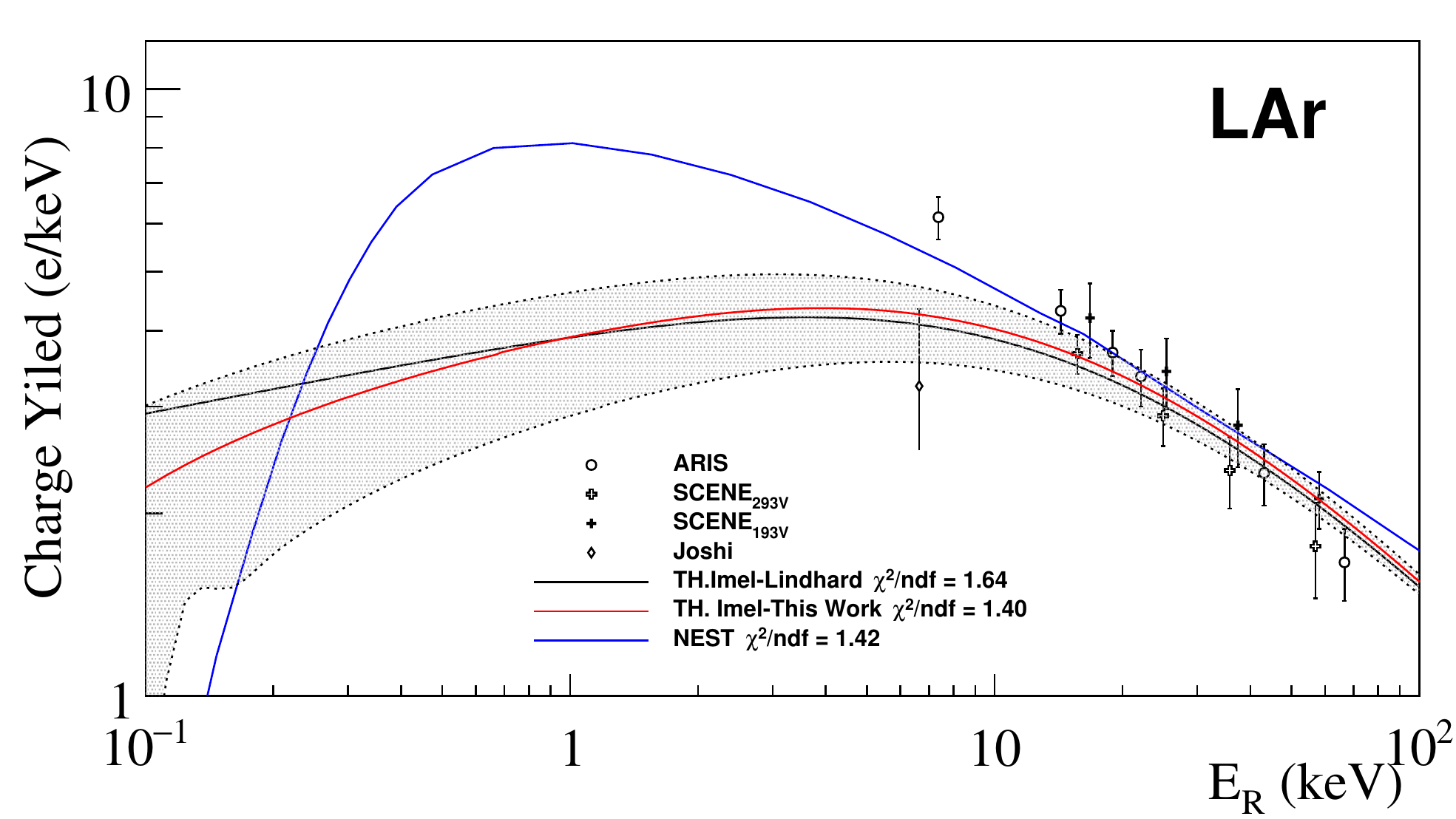}
\caption{Charge yield for LXe and LAr  as a function of the recoil energy. The solid red is the  best fit  using Thomas-Imel box model with the ionization efficiency computed according to \cite{QFUNAM2020}, solid black is Thomas-Imel box model with Lindhard's ionization efficiency and solid blue is NEST curve.}\label{Fig3}
\end{figure}

For total quanta in LXe and LAr  we proceed according to Eq. (\ref{Eq3}), where $f_n=W_s(n_e/E_R+n_\gamma/E_R)$. For simplicity, in the case of LXe we follow ref. [1] and use the subset of available data spanning the energy range from 0.5 to 120.0 keV \cite{NESTData}, removing datasets that overlap in energy, but retaining the general trend in the whole energy range. Adding all the data points does not have a significant impact on our results. For LA we use all currently available points. We ignore the electric field of the measurements, since total quanta,  in principle, independent of this variable. We solve the integro-differential equation for average atomic motion by  the method described in \cite{QFUNAM2020}, varying  the constant binding energy $U$ from 10 to 200 eV and the $k$ factor from 0.083 to 0.2. Minimizing   $\chi^2 =\sum^N_i (D_i-f_n(E_i))^2 /\sigma^2_i$, with $f_n$ for each data point $D_i$ at energy $E_i$, allow us to find the best fit values of $k$ and $U$. For LXe we get $U=30^{+60}_{-16}$ eV and $k=0.090^{+0.013}_{-0.007}$, and for LAr we get $U=20^{+30}_{-10}$ eV and $k=0.095^{+0.013}_{-0.009}$, see Fig.\ref{Fig1}.
For the light and charge yield models we made use of Eq.(\ref{Eq3}) and Eq.(\ref{Eq4}) in addition to the best model for ionization efficiency $f_n$ consistent with errors given by the data points in Fig.(\ref{Fig1}). We implement a $\chi^2$ fit for charge and light  yield  measurements independently, taking as a free parameters $W^*_i$, $\gamma$, $\beta$. For light yield we also take into account the Penning effect by means of the model described in \cite{Penning} that uses an extra free parameter $\eta$. Note that there is a reduction in both light and charge yield for $E_R < 10$ keV, which is more pronounced when using our model instead of the Lindhard formula.  This is because we include the binding energy in the calculation of $f_n$.  The measurements where chosen  for non zero electric fields points around $\approx 300$ V/cm. For LXe we obtained $\gamma=0.013\pm 0.004$, $\beta=0.58\pm 0.12$, $W^*_i=18.2\pm 0.4$ eV and $\eta=0.85\pm0.11$, while for LAr  $\gamma=0.020\pm 0.003$, $\beta=0.935\pm 0.31$, $W^*_i=25.0\pm 1.2$ eV and $\eta=0.49\pm0.10$, see figures \ref{Fig2} and \ref{Fig3}.  
\section{Conclusions}
In this work we present a first-principle study based on the Lindhard integral equation for nuclear recoil ionization efficiency $ f_n$ in LXe and LAr. We show the  dependence of charge and light yield on $ f_n$. The model predicts the turnover of $f_n$ at low energies, already observed in Xe for $E_R<1$ keV. The values obtained from Thomas Imel box model are comparable to other previous studies and the value obtained for  $W^*_i$ energy  is in the expected range for  $W_i$ in liquid  and  gas phases for LXe and LAr. Lindhard's model fails to predict this effect. At higher energies, the model for $f_n$ can be improved  by considering Bohr stripping effects for electronic stopping power. Lindhard integral equation with binding energy is a promising first-principles approach  to study  signal production in noble elements.      
\acknowledgments
This project has received funding from European Union's Horizon 2020 research and innovation programme under grant agreement No. 952480. Also, this research was supported in part by DGAPA-UNAM grants number PAPIIT-IN106322 and PAPIIT-IT100420, and Consejo Nacional de Ciencia y Tecnolog\'ia (CONACYT) through grant CB-2014/240666. 
\newpage
\bibliography{bibliografy.bib}
\bibliographystyle{bibteXJINST}

\end{document}